# Lift augmentation by incorporating bend twist coupled composites in flapping wing

Rahul Kumar [a], Devranjan Samanta [a,*,1] and Srikant S. Padhee [a,*,2]

[a] *Department of Mechanical Engineering, Indian Institute of Technology Ropar Rupnagar-140001, Punjab, India*

*Corresponding author:

[1]E-mail: devranjan.samanta@iitrpr.ac.in

[2]E-mail: sspadhee@iitrpr.ac.in

**Abstract**

Drawing inspiration from the adaptive wing shape of birds in flight, this study introduces a bio-inspired concept for shape adaptation utilizing bend-twist coupling (BTC) in composite laminates. The primary aim of the design optimization is to identify the optimal fibre orientation angles needed to produce the required bending and twisting deformations, which directly contribute to the design's goal of maximizing lift without relying on external mechanisms for twisting. This novel technique increases lift by up to five times compared to a curved bending wing. We have highlighted the vortex dynamics to provide insight into the underlying reasons for such a significant lift increment. In addition, the study presents the Von Mises stress experienced by the wing, offering a comprehensive understanding of the structural behavior. Furthermore, it highlights a significant improvement in efficiency, particularly within the optimal reduced frequency range of 0.25 to 0.4. These findings underscore the potential of this method for future applications in biomimetic drones, micro-air vehicles, and other flapping wing-based systems, ultimately paving the way for new advancements in aerodynamics and structural optimization for next-generation aerial vehicle designs.

*Keywords: Unsteady aerodynamics, Deformable wings, Fiber orientation, Fluid-Structure Interaction*

## 1. Introduction

The evolution of insects and birds to achieve flight through flapping wings has been an intriguing topic for both engineers and biologists [1], [2], [3], [4]. Birds and insects generate both lift and thrust simultaneously through their wings [5], [6]. To achieve the necessary changes in lift and thrust, they adjust the shape of their wings and modify their wing kinematics [2], [6]. Observations show that birds alter their kinematics across different speeds [7] and exhibit wing-flapping behaviours that currently lack a clear mechanistic explanation. In flapping flight, lift and thrust are inherently interconnected, and the way birds flap their wings likely influences the efficiency of generating both forces simultaneously, though not necessarily in the same manner. This dynamic interplay creates an optimum wing kinematics to achieve force balance as efficiently as possible.

Understanding how birds fly and why they flap their wings the way they do provides valuable insights into the evolution of flight, wing structure, and the diverse ways animals navigate the skies [5], [8], [9]. However, much of our knowledge about how wing movements influence flight comes from either simplified models or studies that examine wing motion in real birds. These studies face a significant challenge: birds often adjust multiple aspects of their wing movements simultaneously, such as flapping frequency, stroke amplitude, and angle of attack, making it difficult to isolate the impact of each specific motion on flight performance. As a result, we still lack a comprehensive understanding of how particular wing motions or changes in motion during varying flight conditions affect performance and efficiency [10], [11]. To address this, we need to explore alternative mechanisms birds might use to optimize thrust and lift, such as the automatic adjustment of wing shape to enhance lift force during flapping motion, which could enable optimal performance across a wide range of flow velocities and conditions. This requires research that extends beyond traditional bird studies and delves into innovative approaches to uncover the most efficient strategies for flight.

For MAV applications, there could be benefits in using wings that employ a straightforward kinematic pattern like sinusoidal plunging motion. Although this motion can be easily implemented in practice, pure sinusoidal oscillations are not very efficient at generating lift when using rigid wings. In our previous study [8], we analyzed four different wing configurations: rigid, bending, twisting, and bending-twisting coupled (BTC) wings. We found that compared to the rigid wing subjected to purely sinusoidal motion, the twisting and BTC wings achieved higher lift and propulsive efficiency. Similarly, our study of numerical simulation mimicking the swimming motions of Manta rays showed that while a purely flapping wing generated less lift, variation of the fin length and twisting during the flapping motion significantly enhanced propulsive efficiency [9]. Previous research [8], [9] has demonstrated that twisting in wings is highly effective in attaining the desired aerodynamic performance. However, when it comes to designing the wing to achieve the desired deformation through passive deformation, composite structures are commonly utilized. Composite structures find widespread application across various engineering fields, from marine turbine blades [12], [13] to aircraft wings [14]. This is attributed to their versatility in composite stacking sequences within laminate. This adaptability allows composites to precisely fulfil design requirements for specific applications. Moreover, leveraging the anisotropic mechanical properties of composites enables the realization of coupled behaviours like bend-twist coupling.

Researchers have examined the trade-offs associated with composite coupling effects in structural systems by employing multi-objective optimization. To achieve weight reduction, contemporary aircraft are adopting greater flexibility. Certain highly flexible designs, characterized by large aspect ratios, offer numerous aerodynamic advantages, including decreased induced drag [15] and enhanced aerodynamic efficiency [16], [17], [18], [19], particularly in micro air vehicles (MAVs). Bottaso *et al.* [12] conducted a multilevel finite element optimization constrained by fibre orientation, aiming to design passive BTC composite wind turbine blades that achieve both desired load mitigation and weight reduction. Roeleven [13] utilized a finite element beam formulation and cross-sectional analysis software to optimize BTC wind turbine blades, considering structural constraints such as blade twist, deflection, eigenfrequencies, and material strength. The objective was to minimize blade deflection, optimize blade twist, and manage bend-twist coupling effects. Ozbay [20] demonstrated an optimization approach for extension-twist coupling in composite rotor blades, focusing on enhancing tiltrotor aircraft performance by using a box-beam cross-sectional model to determine coupling distributions. Design variables included cross-sectional geometry and composite layup. DiPalama *et al.* [21] performed optimization of extension-twist coupled composite rotor blades to passively adjust elastic twist distribution according to rotor rotational speed while adhering to material strength constraints under centrifugal and aerodynamic loads.

Numerical methods focused on fluid structure interaction i.e. FSI [22] are crucial for predicting aeroelastic behavior in both the aircraft industry [14] and the biomedical sector [23]. Mian et al. [24] examined the effects of geometric nonlinearity on the static aeroelastic behavior of high aspect ratio wings, finding that their predictions for static deformation and twist were consistent with experimental data from Deman and Dowell [25]. Smith et al. [26] addressed aerodynamic nonlinearity using an Euler solver, revealing that a linear aerodynamic model might yield inaccurate predictions for flutter and divergence speeds. Garcia [27] investigated the aeroelastic behavior of flexible wings under transonic conditions using the Reynolds-averaged Navier-Stokes (RANS) equations and a nonlinear three-dimensional finite element model. They reported that large bending deflections were due to nonlinear torsion-bend coupling effects. Cui and Khoo [28] used FSI simulations to explore the complex aerodynamic responses and stall mechanisms of wind turbine blades under various structural damping levels, showing that nonlinear large deformations and flow field factors can complicate aerodynamic elastic responses. These numerical methods are also applied to studying FSI mechanisms such as the interference effects of vibrating solids on flow fields [29] and the active control of gust responses [30]. Similarly, Han et al. [31] investigated how the position and material of flexible wings affect unsteady flow control and deformation patterns through FSI simulations. This methodology is also relevant to the elastic body FSI issues addressed in this paper [32], [33].

In this study, we employ a fully coupled fluid-structure interaction three-dimensional simulation to investigate the aerodynamic performance of flexible flapping wings. Instead of emulating a complex insect-like stroke, our focus is on elastic wings driven by sinusoidal oscillations. We have investigated the enhancement of aerodynamic efficiency in flapping wings driven by through the utilization of fibre-reinforced composites for wing flexibility. Fiber-reinforced composites offer not only high specific stiffness and strength but also enable customization of directional stiffness as needed [34], [35]. These materials allow precise control over the orientation of fibres in each layer, thereby achieving the desired elastic properties. The numerical model highlights the coupling effects of fibre orientation and material properties on wing deformation and performance, offering opportunities for passive control by adjusting fibre orientations. Specifically, we examine how variations in fibre orientation and laminate layup influence wing twist angles to optimize aerodynamic performance. Notably, the bend-twist coupled (BTC) wing configuration generates lift more than five times greater than bending cases. An optimal laminate configuration is identified, maximizing twist without excessive bending to achieve the desired lift. This requires a carefully selected fibre orientation that facilitates the necessary twisting motion while maintaining controlled bending. Our findings suggest that flexible wings with simplified stroke kinematics can be effectively utilized in the design of efficient flapping Micro Air Vehicles (MAVs). Furthermore, vortex structures such as trailing or leading-edge vortices (TEVs or LEVs) and trailing vortices (TVs) are analyzed using the Q-criterion, revealing mechanisms behind lift and thrust. While earlier studies showed deformable wings outperform rigid ones [4], they didn't account for fluid-structure interaction (FSI), which considers wing elasticity and its impact on twisting angles during flapping motion.

2. Methodology

This study explores the sinusoidal flapping motion of a rectangular wing within the context of fluid-structure interaction. The simplified wing geometry was modelled using ANSYS Workbench Design Modeler CAD software. This study investigates the dynamics of a rectangular wing undergoing sinusoidal flapping motion. The Reynolds numbers ($Re$) are specified as 30,000 (for comparative analysis with published experimental data [36]) and 1.5

× $10^5$ for our main study, based on a free stream velocity ($U_\infty$) of 11m/s and the chord length of the wing. The selected *Re* range is consistent with our previous research [37], where the same value was also used based on free stream velocity. Additionally, similar *Re* have been employed in earlier studies [38] related to low-speed aircraft, making it a suitable choice for the current investigation. To calculate $e = \rho_f U_\infty c/\mu$, we used the following values: the air density, $\rho_f$, is 1.225 kg/m³, the dynamic viscosity and $\mu$, is $1.7894 \times 10^{-5}$ kg/(mis). The Mach number (*Ma*), defined as the ratio of the object's speed ($U_\infty$=11m/s), consistent with our previous research [8], to the speed of sound in the surrounding air (343.2 m/s), is 0.004 for validation and 0.032 for our main study, respectively, confirming that the flow is incompressible.

## 2.1 Governing equations

The aerodynamic characteristics of flapping wing motion are governed by the unsteady, incompressible Navier–Stokes equations, which are solved using the Large Eddy Simulation (LES) approach. In LES, the governing equations are spatially filtered to resolve the large energy-containing turbulent structures while modelling the effects of smaller sub grid-scale (SGS) motions. Assuming an incompressible flow and denoting the filtering operation with an overbar, the filtered Navier–Stokes equations are expressed as:

$$\frac{\partial \bar{u}_i}{\partial x_i} = 0 \quad (1)$$

$$\frac{\partial \bar{u}_i}{\partial t} + \frac{\partial \bar{u}_i \bar{u}_j}{\partial x_j} = -\frac{1}{\rho}\frac{\partial \bar{p}}{\partial x_i} + \frac{\partial}{\partial x_j}\left(\nu \frac{\partial \bar{u}_i}{\partial x_j} - \tau_{ij}^r\right), \quad (2)$$

Here, $\bar{u}_i$ represents the filtered velocity components, $\bar{p}$ is the filtered pressure, $\nu$ is the kinematic viscosity, and $\tau_{ij}^r$ denotes the residual sub grid-scale stress tensor that encapsulates the influence of the unresolved scales. The SGS stress tensor is defined as:

$$\tau_{ij}^r = \overline{u_i u_j} - \bar{u}_i \bar{u}_j \quad (3)$$

$$\tau_{ij}^r = -2\nu_t \bar{S}_{ij} + \frac{1}{3}\delta_{ij}\tau_{kk} \quad (4)$$

The Kronecker delta, $\delta_{ij}$, equals 1 when $i$ is equal to $j$, and 0 otherwise. Here, $\nu_t$ represents the subgrid-scale (SGS) eddy viscosity, while $\bar{S}_{ij}$ denotes the rate-of-strain tensor corresponding to the resolved scale, which is defined by:

$$\bar{S}_{ij} = \frac{1}{2}\left(\frac{\partial \bar{u}_i}{\partial x_j} + \frac{\partial \bar{u}_j}{\partial x_i}\right) \quad (5)$$

The mixed time-scale (MTS) model, as presented in [39], has been employed in this study. One of the key benefits of this model is its inherent ability to reduce eddy viscosity to zero near the wall, thereby eliminating the need for empirical constants or damping functions when dealing with wall-bounded flows. Additionally, the model has shown reliable performance in simulating transitional flows over oscillating wings at similar Reynolds numbers, as demonstrated in [40].

## 2.2 Aerodynamic performance calculation

An object or creature in flight, whether a bird, bat, insect, or airplane, experiences four primary forces: lift, thrust, drag, and gravity. Birds generate lift and thrust through wing flapping, which involves a complex and dynamic three-dimensional motion that changes with each wing position [9], [37], [41]. This flapping action includes two phases: the downstroke or power stroke, which produces most of the thrust, and the upstroke or recovery stroke, which creates varying amounts of thrust depending on the wing's shape [42]. The aerodynamic force on the wing, influenced largely by stroke velocity, comprises both lift and thrust [42]. The key aerodynamic parameters for wing strokes are the lift coefficient ($C_L$) and the thrust coefficient ($C_T$), which are defined as follows:

$$C_L = \frac{F_L}{\frac{1}{2}\rho_f U_\infty^2 A} \quad (6)$$

$$C_T = \frac{F_T}{\frac{1}{2}\rho_f U_\infty^2 A} \quad (7)$$

$F_L$ and $F_T$ represent the lift and drag forces, respectively; $A$ is the projected area.

The aerodynamic power can be expressed using the following equation [14]:

$$dP_{aero}(t) = \left(p(t) \cdot \overrightarrow{dA}(t) + \vec{\tau}(t) \cdot dA\right) \cdot \vec{v}(t) \quad (8)$$

Where, $p, \overrightarrow{dA}, \vec{\tau}, \vec{v}$ represent the pressure, area vector, viscous force velocity and elemental area on the element respectively.

The total aerodynamic power, $P_{aero}$, is obtained by integrating $dP_{aero}$ over the closed surface area $A$ of the wing exposed to the fluid:

$$P_{aero}(t) = \oint_A dP_{aero}(t) \quad (9)$$

Next, we compute the mechanical input power, $P_{mech}$. This is obtained by multiplying the mechanical moment $M_Y$ by the angular velocity $\dot{\delta}$, as given by the following equation:

$$P_{mech}(t) = M_Y \cdot \dot{\delta}(t) \quad (10)$$

The power coefficient, $C_P$, is then calculated using the formula:

$$C_P(t) = \frac{P}{\frac{1}{2}\rho_f U_\infty^3 A} \quad (11)$$

Finally, we define the aerodynamic efficiency of the flapping wing as follows:

$$\eta = \frac{\bar{C}_L}{\bar{C}_P} \quad (12)$$

## 2.3 Computational details

The simulations were conducted using the ANSYS FLUENT commercial finite-volume solver, with the Large Eddy Simulation (LES) model to simulate the flow field. The two-way coupled Fluid-Structure Interaction (FSI) model was implemented by integrating fluid dynamics with structural mechanics at each time step [43]. In a two-way coupled FSI approach, the fluid flow affects the deformation of the structure, while the resulting structural deformation, in turn, modifies the fluid domain. This mutual interaction is resolved through an iterative process that continues until convergence is achieved for both the fluid and structural fields at each time step. A constant velocity of 11 m/s was imposed at the domain entrance. A pressure boundary condition was applied at the exit. No-slip condition was set at the wall. The numerical analysis in this study utilizes a solver that incorporates unsteady simulations and dynamic mesh capabilities. Tetrahedral cells were used for the fluid domain, while quadrilateral cells were employed for the structural domain, with the total number of elements in the fluid domain being 3,284,585, providing sufficient spatial resolution to capture key flow structures [44], [45]. For the pressure-velocity coupling, the Pressure-Implicit with Splitting of Operators (PISO) scheme is utilized [46]. Specifically, the pressure field is discretized using a second-order scheme, while momentum utilizes a second-order upwind scheme. Spatial discretization is achieved using a least-squares cell-based method. Turbulent kinetic energy (TKE) is discretized using first-order upwind schemes. TKE represents the energy stored in turbulent eddies and is defined as the average kinetic energy per unit mass due to turbulent fluctuations. Furthermore, numerical stability is ensured by enforcing the Courant–Friedrichs–Lewy (CFL) condition with a Courant number maintained below unity as prescribed by the governing equation.

$$C_n = \frac{U_\infty \Delta t}{\Delta x} \leq 1 \quad (13)$$

The Courant number $C_n$ is defined as the ratio of the velocity magnitude $U_\infty$ multiplied by the time step $\Delta t$, to the mesh node interval $\Delta x$. In this simulation, a fixed time step of 0.005s seconds is employed, with a convergence criterion set at residual values of $10^{-6}$.

3. **Wing geometry, kinematics and material properties**

The rectangular wing (600 mm x 200 mm x 1.2 mm) experiences flapping motion via a revolute joint at its root, with rotation of the wing occurring about the y-axis through this revolute joint. The dimensions of the rectangular wing have been selected based on the previous study [8], in which the same aspect ratio was used. Consequently, the geometry was kept fixed in this study, and the other parameters were varied for analysis. The flapping motion of the wing is defined by:

$$\delta(t) = \delta_0 + \delta_1 \sin(\omega t) \quad (14)$$

This sinusoidal flapping motion is also defined in our previous paper for both 2D [47] and 3D [8] cases and the same equation was also employed in the study by Ghommem *et al.* [48]. Additionally, the previous study suggests using this equation for the flapping motion [49], [50]. where, $\delta_0, \delta_1$ and $\omega$ represents the initial inclination, angular amplitude, and angular frequency of flapping, respectively. *t* represents time in sec and $T_p$ denotes the period of a flapping cycle. In this simulation, a fixed $\delta_1$ of 30° and $\delta_0$ of 0° are employed, with the angle of attack ($\alpha$) maintained constant at 10°, at $\alpha$ =10° the BTC wing has highest lift coefficient (refer Fig. 6). The kinematic parameters used in this study are identical to those employed in our previous research [35]. The influence of unsteady flow on the aerodynamic efficiency of flapping wings is investigated across different reduced frequencies. Reduced frequency is a dimensionless parameter crucial in unsteady aerodynamics, defined as $k = \omega(c/2)/U_\infty$.

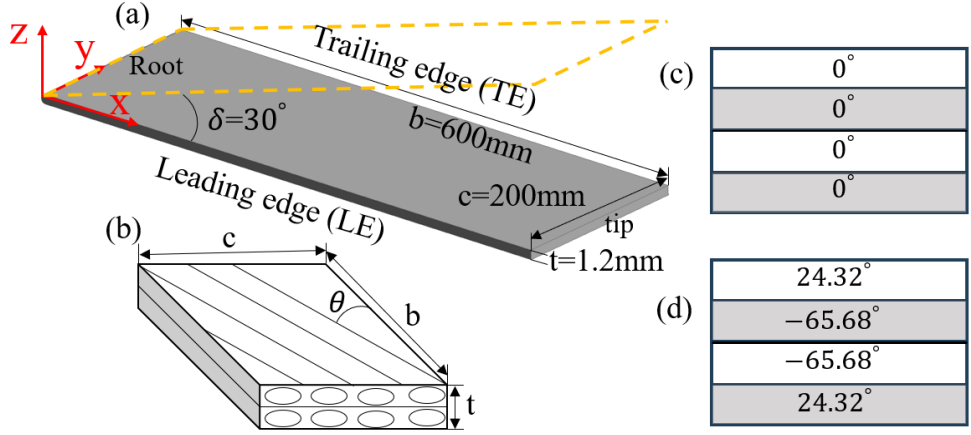

Fig. 1. (a) Solid wing geometry, where the gray solid part represents the wing in its mid-position, and the yellow dotted line indicates the wing's position at the uppermost extent during flapping. (b) Fiber orientation of the laminate, with $\theta$ representing the angle of fibre orientation. (c) Unidirectional laminate (bending wing). (d) Fiber orientation of BTC wing laminate.

We have leveraged the anisotropic mechanical properties of composites to achieve the bending-twisting coupling (BTC) mode for optimum aerodynamic performance. This approach involves selecting anisotropic material properties and arranging differential ply layering in a stacking sequence. In Figure 1b, we can observe that the fibre orientation, denoted as $\theta$, can be varied for different layers to achieve the optimal bend-twist coupling value. The primary comparison in this paper is between a unidirectional laminate [0/0/0/0] and a fibre orientation represented by $[\{24.32\}_2/\{-65.68\}_2]$ the two extremums of the bend twist coupling [35], and illustrated in Figures 1c and 1d, respectively. This comparison will be discussed further in the subsequent sections of the paper. Table 1 provides the detailed material properties for IM7/8552 [51]. This method eliminates the necessity of external mechanisms for inducing twists, relying solely on the anisotropic properties of the materials and the optimized stacking sequence.

Table 1. Key material properties of IM7/8552 [51].

| $E_1$ (GPa) | $E_2$ (GPa) | $E_3$ (GPa) | $\lambda_{12}$ | $\lambda_{13}$ | $\lambda_{23}$ | $G_{12}$ (GPa) | $G_{13}$ (GPa) | $G_{23}$ (GPa) | $\rho_s (g/cm^3)$ |
|---|---|---|---|---|---|---|---|---|---|
| 161.0 | 11.38 | 11.38 | 0.32 | 0.32 | 0.45 | 5.17 | 5.17 | 3.92 | 1.57 |

Here E and G represent elastic modulus and shear modulus respectively. The subscripts 1,2 and 3 represent x, y and z respectively. Here λ denotes Poisson's ratio.

## 4. Validation

To validate the force measurement system, experiments were conducted on a NACA0012 airfoil with a 100mm chord and 400mm span. The airfoil oscillated with a non-dimensional constant amplitude ($h = 0.175$) between two end plates. Tests were performed at $Re = 30{,}000$ and a reduced frequency ($k$) of 1.82. The temporal variations of the lift coefficient ($C_L$) and thrust coefficient ($C_T$) are depicted in Fig. 2(a) and 2(b), respectively. For accuracy assessment, results were compared with those obtained from the discrete vortex method (DVM), as detailed in prior studies [8], [9]. In the case of DVM simulations, the wing experiences active deformation, implying that the prescribed deformation is imposed upon the wing. In the current simulation, the deformation is passive, implying that the wing deformation is also affected due to the pressure distribution at each time step. To ensure a proper comparison, the same deformation should be applied in the DVM case, while also maintaining the kinematic parameters. The values of $C_L$ in current simulations were found to deviate by less than 6% when compared with both Gordnier *et al.* [37] and DVM results shown in Fig. 2(a). In Fig. 2(b), the temporal variation of the thrust coefficient $C_T$ was compared, revealing a maximum deviation of less than 4% with both the experimental [52] and DVM results. These discrepancies can be attributed to mechanical effects in experimental results, such as linkages, which are not accounted for in simulations. Additionally, while the DVM employs active morphing, our simulations involve passive morphing, posing challenges in replicating identical deformations. Furthermore, the DVM is an inviscid model that excludes considerations of viscous effects and LEV, both of which are included in our current simulations.

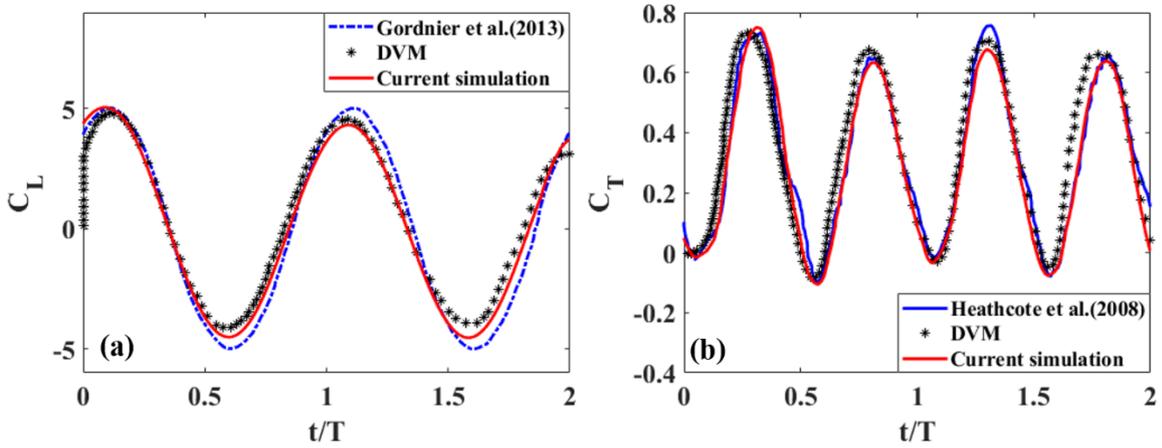

Fig. 2. (a) Comparison of lift for a flapping wing with results obtained by Gordnier *et al.* [37] and DVM. (b) Comparison of thrust with experimental results [52] and DVM [8].

## 5. Results and discussions

In the realm of aerospace engineering, understanding the intricate interplay between structural deformations and aerodynamic forces is pivotal. Our study employs two-way Fluid-Structure Interaction (FSI) to delve into the behavior of flexible flapping wings under varying reduced frequencies ($k$). The solid solver meticulously tracks wing deformations, while the fluid solver analyzes crucial aerodynamic parameters such as $C_L$, $C_T$, $C_P$ and vorticity, and wing surface interactions with fluid. Through our simulations, we unveil how these dynamic interactions shape both the structural integrity and aerodynamic performance of the wing. This research sheds light on fundamental principles governing flight dynamics, with implications ranging from bio-inspired robotics to advanced aircraft design.

### 5.1 Structural Analysis

In this section, we discuss how different degrees of fibre orientation in composite materials lead to varying levels of bending deflection and twisting under different loads. However, accurately determining these deflections and twists through parametric studies can be challenging, often requiring numerous iterations to achieve optimal designs. In this study, we conducted an extensive literature review to explore how maximum twist can be achieved solely by varying fibre orientations and to understand how much deflection and twist are necessary to maximize aerodynamic performance. We have conducted simulations with different fibre orientations to determine the sufficient levels of deflection and twist. Table 2 below summarizes the results for each stacking sequence under an aerodynamic load at an angle of attack of 10°. This angle of attack was selected as highest $C_L$ was achieved at $10^o$ by the BTC wing (refer Fig. 6). This stacking sequence is based on previous studies in which researchers reported that, in these stacking sequences, both bending and twisting of the wing occurs [34], [35], [53]. We have verified all the stacking sequences and analyzed them to determine which stacking sequence results in the maximum twist. When the fibre orientation ($\theta$) is kept at 0° for all plies, bending occurs with significantly low twisting.. However, as soon as the value of $\theta$ is changed, twisting begins to occur in addition to bending in the wing. Throughout the study, we maintained a fixed load condition with a freestream velocity of 11 m/s, consistent with our previous research [8], and optimized the thickness of the wing to achieve maximum deflection without failure, resulting in an optimal thickness of 1.2 mm. In Table 2, detailed information is provided regarding the maximum twisting angles observed in the wings at specific times when twisting is at its peak. These angles are recorded for different stacking sequences and correspond to the bending angles of the wings at that moment. The maximum twist is achieved with the stacking sequence $[\{24.32\}_2/\{-65.68\}_2]$, which has also been reported in previous studies as the sequence that maximizes twist [35].

Table 2. Effect of composite layups on the structural behavior of the wing.

| Fiber orientation (°) | Bending, $def$ (mm) | Twisting angle, $\emptyset$ (°) | Von Mises Stresses (MPa) |
|---|---|---|---|
| [0/0/0/0] | 66.12 | 1.22 | 157 |
| [45/45/45/45] | 76.24 | 9.09 | 159 |
| [24.32/-65.68] | 79.55 | 14.53 | 151 |
| $[\{24.32\}_2/\{-65.68\}_2]$ | 84.636 | 17.20 | 143 |
| $[\{24.32\}_3/\{65.68\}_3]$ | 72.64 | 10.70 | 149 |
| [45/90/90/135] | 69.24 | 7.45 | 141 |

Figure 3 (a) illustrates the clear visualization of the variations in $\delta$, deflection ($def$) and twisting angle ($\emptyset$) over time for a bending wing with a fiber orientation of [0/0/0/0], compared to a BTC wing with a fibre orientation of $[\{24.32\}_2/\{-65.68\}_2]$. The deflection ($def$) is observed as the deviation of the structure from its original position, caused by applied forces. Twisting is depicted as the rotational displacement around the longitudinal axis, resulting in an angular distortion. For the BTC wing, the maximum twisting angle ($\emptyset$) of 17.20 degrees and the maximum bending ($def$) of 84.636 mm are achieved at the midpoint of the stroke ($t/T\sim0.5$). In contrast, for the bending wing, the twisting angle is negligible, with the maximum bending of 66.12 mm occurring at $t/T\sim0.35$. Figure 3 (b) illustrates the relationship between reduced frequency and the maximum twist angle ($\emptyset$). It demonstrates that $\emptyset$ increases as the frequency rises. As the frequency increases, the wing experiences greater forces, leading to a higher maximum twist angle.

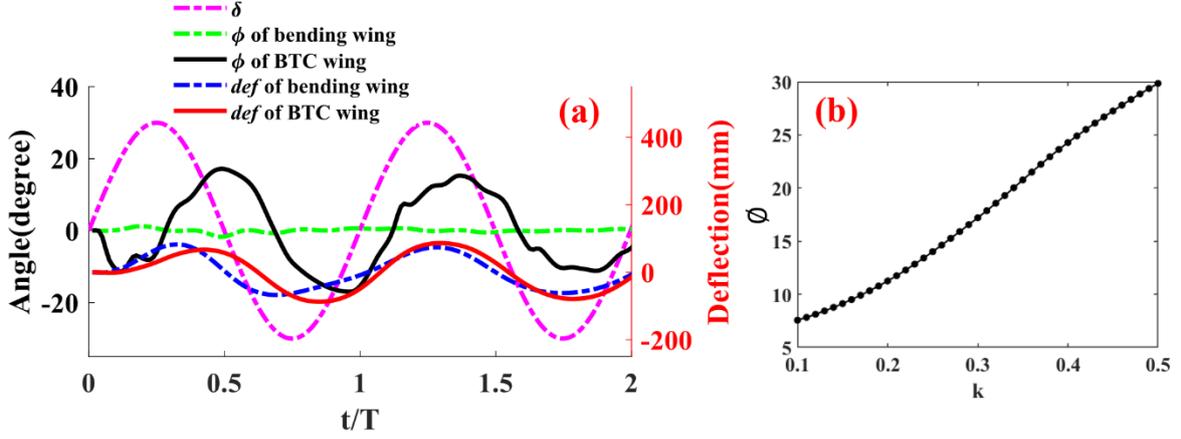

Fig. 3. (a) Illustrates the key mechanical deformation aspects, including flapping, twisting and bending. The deformation effects are analyzed for two different fibre orientations: BTC wing fibre orientation [$\{24.32\}_2/\{-65.68\}_2$] and bending wing fibre orientation [0/0/0/0], (b) The effect of reduced frequency on the maximum twisting angle ($\emptyset$) of BTC wing.

The data transfer from Fluent to Transient Structural is accomplished through Fluid-Structure System Coupling [43]. The pressure distribution obtained from the CFD analysis of the wing at each time step is used in the transient structural analysis to calculate the Von Mises stress ($\sigma_{VM}$) on the wing, as shown in Figure 4. Von Mises stress is used in FEA to predict material yielding under complex, multi-axial loading conditions. It transforms the multi-directional stress components into a single scalar value, which can be directly compared with the material's yield strength to assess structural safety and performance [54]. The $\sigma_{VM}$, is calculated using the following expression:

$$\sigma_{VM} = \sqrt{\frac{1}{2}\left[(\sigma_x - \sigma_y)^2 + (\sigma_y - \sigma_z)^2 + (\sigma_z - \sigma_x)^2 + 6(\tau_{xy}^2 + \tau_{yz}^2 + \tau_{zx}^2)\right]}, \quad (15)$$

Where, $\sigma_x, \sigma_y, \sigma_z$ are the normal stress components in the x, y, and z directions, respectively and $\tau_{xy}, \tau_{yz}, \tau_{zx}$ are the shear stress components in the respective planes.

This equation consolidates the three-dimensional stress state into a single scalar value, $\sigma_{VM}$, which is compared with the material's yield strength $\sigma_y$ to assess failure risk. For the present study, the maximum Von Mises stresses for the bending and BTC wings are found to be 157 MPa and 143 MPa, respectively. Given that the yield strength of the composite material is $\sigma_y$ = 700MPa [55], both configurations operate within safe stress limits. To ensure structural integrity during flight, it is required that $\sigma_{VM} \leq \sigma_y$. The composite layups [0/0/0/0] and [$\{24.32\}_2/\{-65.68\}_2$] demonstrate safety factors of 4.57 and 4.90, respectively.

Thus, for the given geometry and material even under the highest load conditions, the stresses experienced by the composite wing remain well below the yield stress limit. As a result, the composite wing performs efficiently and maintains its structural integrity. Furthermore, the magnitude of the Von Mises stress in a wing undergoing bending (Fig. 4a) is higher compared to that in a wing undergoing twisting (Fig. 4b). In a bending wing, the stress is more evenly distributed, particularly at the root, because the entire wing experiences bending forces uniformly. Conversely, in a BTC wing, the trailing edge bends more significantly on one side, leading to a concentration of stress at specific points due to the uneven distribution of forces. Consequently, the higher stress experienced by the bending wing suggests that, from a structural perspective, the BTC wing configuration is more structurally stronger in this scenario.

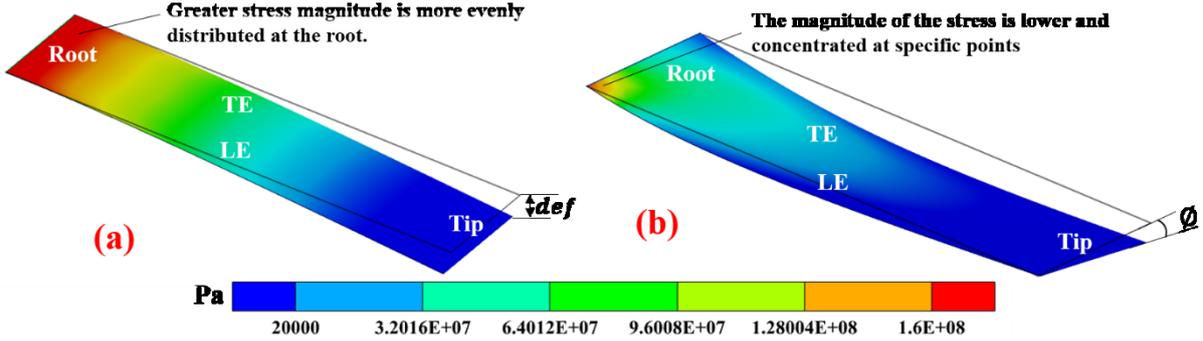

Fig. 4. Von Mises stresses on the wing at $t/T\sim0.5$ for $k$=0.3 (a) Bending case with fibre orientation [0/0/0/0] (b) BTC wing with fibre orientation $[\{24.32\}_2/\{-65.68\}_2]$. The terms $'def'$ and $'\emptyset'$ refer to the deflection and twisting of the wing, respectively.

## 5.2 Lift and drag force

Figures 5(a) and 5(b) illustrate the temporal variation of the coefficient of lift ($C_L$), and the thrust coefficient ($C_T$) over 10 cycles of flapping motion for the bending and BTC modes, with a reduced frequency $k$= 0.3. According to previous studies [56], [57], [58], the majority of the net lift is generated during the downstroke phase of the flapping cycle, while the upstroke phase produces relatively little lift. During the upstroke, birds perform actions such as pulling in their wingtips or increasing wing sweep [59], which reduces span, drag, and power requirements, thereby enhancing efficiency. The temporal lift plot for the bio-inspired BTC wing and the flexible wing reveals distinct aerodynamic characteristics over multiple flapping cycles. The BTC wing demonstrates higher and more consistent peak lift values, indicative of a stable and efficient lift generation mechanism. In contrast, the flexible wing exhibits lower peak lift values with pronounced fluctuations. We can also observe that there is a noticeable difference in the lift peaks between the two wings. In the BTC wing, not only is there a lift peak at the mid-position of the downstroke, but there is also a small lift peak occurring before the completion of one wing cycle.

To assess the aerodynamic performance of the wing, we conducted a detailed investigation to quantify the lift forces and compare them with the wing's weight. Our primary focus lies in determining the average forces, specifically categorizing lift and thrust in the vertical and horizontal directions. Notably, the instantaneous forces achieve a near-periodic state within a few wing strokes. This rapid transition occurs largely due to minimal interference from vortices generated in preceding cycles, as illustrated in Figure 9. The swift attainment of a periodic state theoretically facilitates the potential for a wing to initiate lift within a limited number of strokes. In analyzing the lift generated by the wing, we can estimate its effectiveness in supporting the wing's weight for flight. The mass of the wing is estimated by multiplying wing volume with the density of the material (1.57 g/cm³). In the case of a bending wing, the lift force is measured at 1.39 N. Conversely, for a BTC wing, the lift force is 6.9 N. Given that the weight of the wing is 2.21 N, the lift produced by the bending wing is inadequate to support the wing for sustained flight. On the other hand, the lift generated by the BTC wing is sufficient to support the wing's weight under the same parameters.

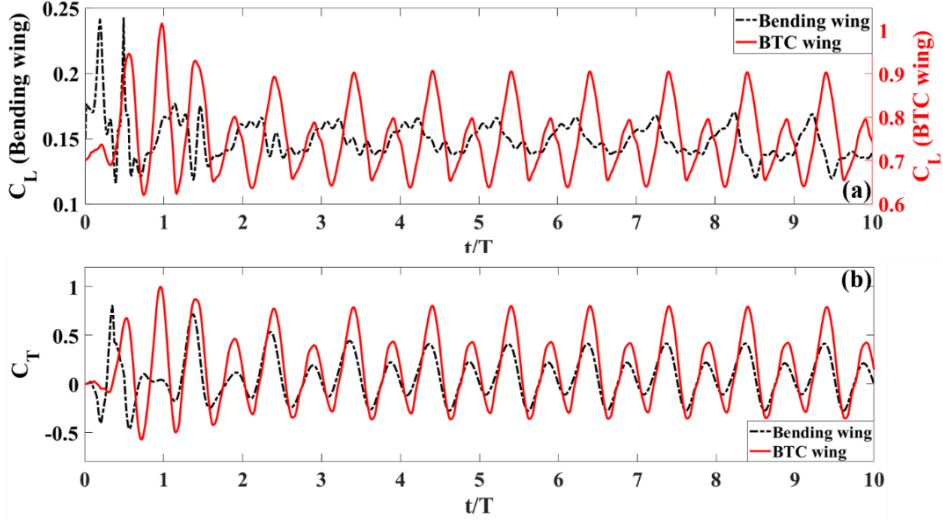

Fig. 5. (a) Lift and (b) thrust coefficient as a function of time over 10 cycles; $k$=0.3.

Instantaneous thrust coefficient curves for the case with $k$=0.3 are shown in Fig. 5(b). Two distinct peaks in thrust are observed during one cycle, which is consistent with our previous results [8] and also aligns with findings from earlier studies [52]. It can be observed that the thrust coefficient for the BTC wing is higher than that for the bending wing, indicating that incorporating a degree of twisting enhances the thrust coefficient. Specifically, the BTC wing achieves a 77% increase in average thrust (averaged over a flapping cycle) compared to the bending wing.

### 5.3 Effect of angle of attack (α) on $\bar{C}_L$

The estimation of the averaged lift coefficient ($\bar{C}_L$) at several angles of attack (α) is crucial for safe operation and optimum aerodynamic performance. Figure 6 illustrates how the α affects the lift for both the bending and BTC wing configurations. As the α increases, the pressure differential between the upper and lower surfaces intensifies, enhancing lift to a critical point. In a BTC wing, variations in α across the span can lead to differential lift, resulting in increased twisting of the wing and allowing it to adapt to changing flight conditions [48]. The bending wing achieved its highest lift coefficient at an α of 15°, while the BTC wing had the highest lift coefficient at an α of 10°, representing 4.75 times increase compared to the bending wing's highest lift coefficient at an α of 15°.

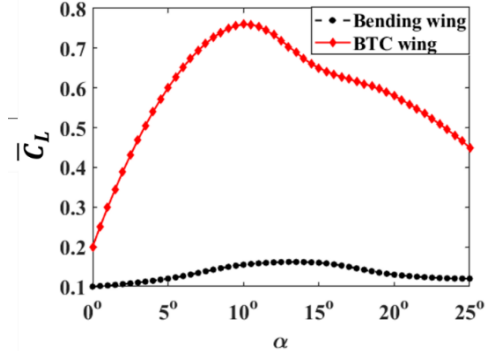

Fig. 6. The effect of angle of attack ($\alpha$) on the averaged lift coefficient ($\bar{C}_L$ was analyzed based on results from 10 cycles of flapping motion).

**5.4 Effect of frequency on aerodynamic performance**

Figure 7 highlights the influence of reduced frequency $k$ on the averaged lift coefficient ($\bar{C}_L$) for bending and BTC wings. The bending wing displays a monotonically increasing linear trend of lift coefficient with increasing frequency, indicating a proportional relationship between the flapping frequency and lift generation. This linearity suggests that as the wing flaps faster, the aerodynamic forces increase consistently, contributing to a higher lift. Conversely, the BTC wing exhibits a nonlinear relationship between lift coefficient and reduced frequency. The lift coefficient for the BTC wing increases rapidly at lower frequencies, reaches a peak at around $k$=0.3 and then starts to decrease with an increase in $k$. This non-monotonic behavior can be attributed to the complex nonlinear interplay between bending and torsional deformations in the wing structure. The initial rapid increase in lift coefficient with frequency indicates that the coupled deformation mechanisms significantly enhance lift at lower frequencies. However, as the frequency continues to increase, the twisting angle also increases (see Fig. 3b), which results in flow separation. This phenomenon subsequently leads to a reduction in lift [60].

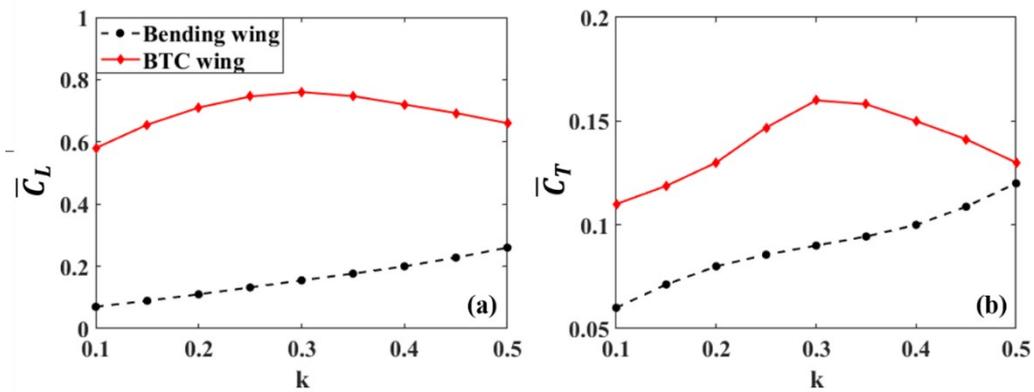

Fig. 7. The effect of reduced frequency on the averaged lift coefficient ($\bar{C}_L$) and thrust coefficient ($\bar{C}_T$) was analyzed based on results from 10 cycles of flapping motion.

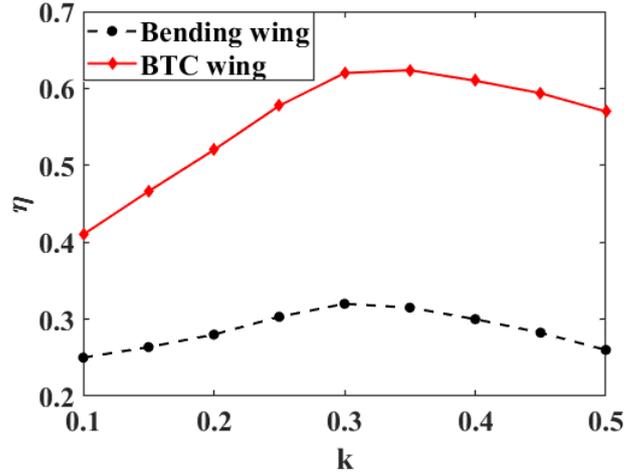

Fig. 8. The effect of reduced frequency on $\eta$ is analyzed based on results from 10 cycles of flapping motion.

Fig. 8 demonstrates the impact of reduced frequency on efficiency for both bending and BTC wings. The efficiency of a wing is influenced by various factors such as the wing's geometry, flow conditions, and operating conditions, with reduced frequency ($k$) being a key factor. The efficiency of a wing increases significantly with reduced frequency up to a moderate range, typically between 0.25 and 0.4, where it reaches a maximum value. Beyond this range, however, efficiency begins to decrease monotonically as reduced frequency continues to increase. This decrease in efficiency at higher reduced frequencies can be attributed to increased drag forces due to greater inertial forces from the rapid flapping wing motions and higher air resistance, adverse flow conditions such as flow separation and vortex shedding [8], [14], [52].

### 5.5 Vortex structure

To provide insight into wing aerodynamics, we present a series of snapshots in Fig. 9 that illustrate the spanwise vorticity contours at a position halfway along the wing's span (0.5$b$) at various times throughout the flapping cycle. At $t/T$ =1.25, there is minimal difference observed between the Leading-Edge Vortex (LEV) and Trailing-Edge Vortex (TEV) of the bending and BTC wings. At $t/T$=1.5, the LEV that has developed interacts with the leading edge of the wing. Since this LEV rotates in a clockwise direction, it increases the effective relative velocity at the leading edge of the wing. This heightened relative velocity generates additional aerodynamic forces on the wing, resulting in a greater lift. Additionally, the increased velocity caused by the LEV enhances the effective angle of attack, which further contributes to the wing's lift production. As a result, the overall aerodynamic performance of the wing is improved [61]. This interaction facilitates the rapid formation of a new LEV on the upper surface of the wing. On the lower surface of the wing, the influence of the anticlockwise vortex is more pronounced in the case of the BTC wing configuration compared to the bending wing configuration. The stronger anticlockwise vortex on the lower surface of the BTC wing contributes to higher lift production, while in the bending wing case, the influence of the anticlockwise vortex on the lower surface is diminished, resulting in a reduction in lift. Du and Sun [62] also observed that the aerodynamic force on the wing could be greater because the LEV is more concentrated and closer to the wing's surface. Additionally, Dickinson *et al.* [63] reported that a smaller LEV tends to be more stable, as the fluid can more easily sustain reattachment for a longer duration, which is beneficial for enhancing lift force. In the case of BTC wing, the orientation of the wing's fibres may lead to a twisting effect, aligning the wing's surface approximately normal to the wind direction. This alignment can enhance the aerodynamic forces experienced by the wing during the downstroke. At $t/T$=1.75, the TEV undergoes partial shedding in the bending-wing scenario. This breakdown of the TEV results in the formation of a small clockwise vortex at the tip of the trailing edge, which alters the momentum of the wing's motion and causes a sudden drop in lift. In contrast, the smooth formation of the TEV in the BTC-wing case results in a gradual increase in lift. At the end of the flapping cycle i.e. $t/T$=2, the leading edge of the BTC wing undergoes a downward displacement. Additionally, the influence of the LEV on the lower surface is greater in the BTC-wing case compared to the bending-wing case. These observations are attributed to elastic recovery and result in an enhanced lift performance for the BTC wing relative to the bending wing [61], [64].

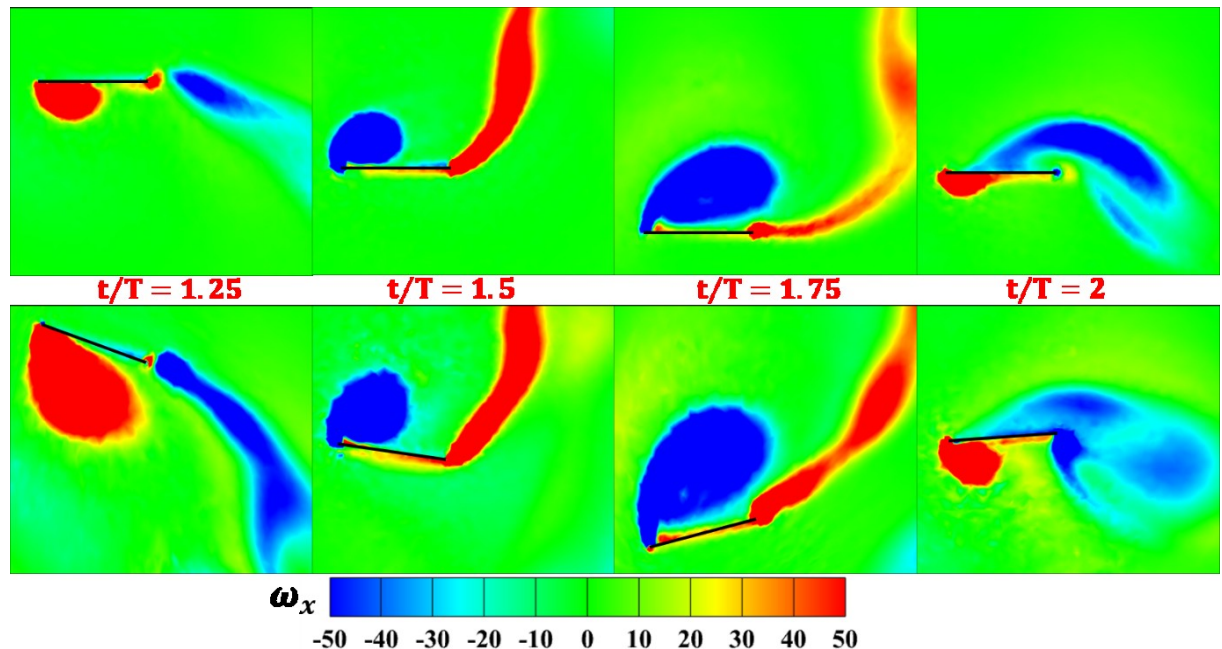

Fig. 9. Spanwise vorticity contours at a position halfway along the wing's span (0.5$b$) at various times throughout the flapping cycle. The first row shows the vorticity contours for the bending wing, while the second row displays the vorticity contours for the BTC wing. Blue indicates clockwise vorticity, and red indicates anticlockwise vorticity. The second flapping cycle is presented to avoid the initial transient effects.

In the study of flapping wing aerodynamics, the flow dynamics are highly intricate, with vortices representing a fundamental and complex flow feature. Understanding these phenomena necessitates effective methods for vortex identification and visualization The temporal evolution of vortex structures such as Trailing-Edge Vortices (TEV), Leading-Edge Vortices (LEV), and Tip Vortices (TV) for both bending mode and BTC wings are depicted in Fig. 10, using the Q-criterion with a value of Q = 30,000 [14] (Multimedia available online). The Q-criterion is a well-established technique for vortex identification that leverages the second invariant of the velocity gradient ($\nabla v$). This tensor's characteristic equation can be expressed as follows [65]:

$$\lambda^3 + P\lambda^2 + Q\lambda + R = 0 \qquad (16)$$

$P$, $Q$, and $R$ denote the invariants of the velocity gradient tensor. In this context, the Q-criterion focuses on the second invariant , $Q$ in eq-16, to distinguish vortices from other flow structures [66]. This method is particularly well-suited for the analysis of vortices in incompressible fluids and has been effectively applied in various numerical simulation studies to examine coherent vortices. The Q-criterion is mathematically defined by:

$$Q = \frac{1}{2}[|\Omega|^2 - |S|^2] > 0 \quad (17)$$

Where,

$$\Omega = \frac{1}{2}[\nabla v - (\nabla v)^T] \quad (18)$$

$$S = \frac{1}{2}[\nabla v + (\nabla v)^T] \quad (19)$$

Where $S$ means the strain rate tensor, $\Omega$ means the rotation rate tensor.

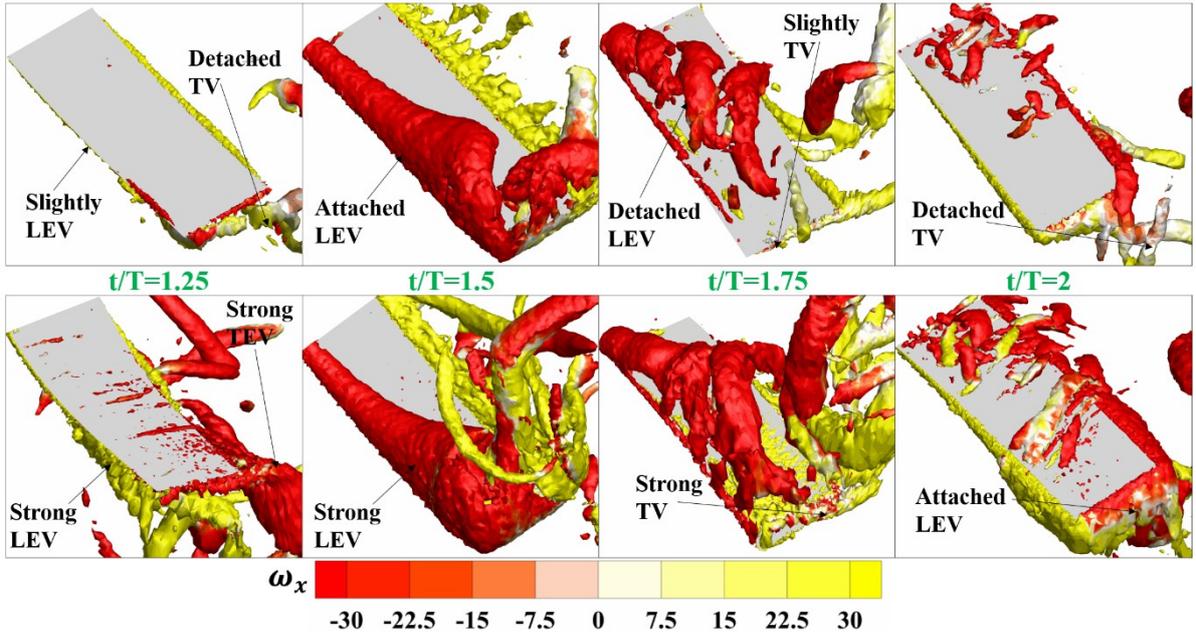

Fig. 10: Application of the Q-criterion for identifying vortex structures for aerodynamic analysis of a wing. The first row of images displays the bending wing, while the second row shows the BTC wing. The second flapping cycle is presented to avoid the initial transient effects. (Multimedia are available online for both bending wing (bending.avi) and the BTC wing (btc.avi)).

Vortical structures such as LEVs, TEVs, and TVs critically influence aerodynamic performance with LEVs generate low-pressure regions (refer Fig. SI-I) that enhance lift, TEVs affect wake dynamics, and TVs create a low-pressure zone near the wingtip (refer Fig. SI-I), further contributing to lift improvement [67], [68]. At $t/T=1.25$, the bending wing exhibits a mix of detached and slightly attached states, indicating partial shedding or weak coherence in the LEVs, while the TVs are fully detached, suggesting complete shedding or breakdown, which prevents the formation of a strong vortex at *t/T*=1.5. In contrast, the BTC wing demonstrates strong, coherent LEVs and TEVs, maintaining a well-formed, high-intensity vortex that enhances vertical force at *t/T*=1.5 (refer fig. 5 (a)). The stable vortices generated by flexible wing flapping contribute significantly to lift augmentation [67]. However, for the bending wing, the LEVs and TEVs begin detaching at *t/T*=1.75 and fully shed by the end of the flapping cycle, leading to inconsistent lift generation and higher energy dissipation due to flow separation, whereas the BTC wing maintains stable vortex attachment. The BTC wing's ability to sustain robust vortical structures highlights the importance of vortex management in wing design, particularly for applications requiring high manoeuvrability and energy efficiency, such as bio-inspired drones or advanced aircraft.

## 6. Conclusion

The present study extensively investigates the numerical simulation of a rectangular flapping wing within the framework of fluid-structure interaction. Following are the few critical observations:

- The use of composite materials for passive deformation allowed effective wing bending twisting coupling. An asymmetric $[\{24.32\}_2/\{-65.68\}_2]$ laminate was used to achieve 17.20° twist and 84.636 mm deflection, replacing external mechanisms with tailored ply orientations for aerodynamic efficiency.
- The study confirms that the Von Mises stress experienced by the composite wing is about five times lower than the yield stress, ensuring that the wing maintains its structural integrity under operational conditions.
- The comparison between bending wings and BTC wings reveals a substantial performance advantage for BTC wings. The increase in lift by 5 times and the 77% improvement in thrust underscore the superior aerodynamic efficiency of BTC wings.
- The use of the Q-criterion to analyze vortex structures reveals that BTC wings generate stronger and more consistently attached vortices. This characteristic is a key factor in the improved aerodynamic performance of BTC wings.
- The study identifies that optimal aerodynamic performance is achieved within the reduced frequency range of *k* = 0.25-0.4. This insight is crucial for designing flapping wing mechanisms to maximize efficiency.

We believe that these findings provide valuable insights for design of flapping based MAVs, emphasizing the benefits of BTC configurations and the importance of material and frequency for enhanced performance. Further it would be interesting to explore the possibility of tailor-ability of such wings for augmenting the lift only or thrust only or both as in this case.

**Supplementary Material**
See supplementary material for the figures of (1) Pressure distribution contours at various times throughout the flapping cycle; not discussed in main text.

**Acknowledgments:**
The authors would like to thank the funding agency' Aeronautics Research and Development Board (ARDB) under the Defence Research and Development Organization (DRDO), Government of India, for the support by Research Grant No. ARDB/01/10312/M/I.

**Data Availability statement:**

The data that support the findings of this study are available from the corresponding author upon reasonable request.

**Supplementary section**

# Lift augmentation by incorporating bend twist coupled composites in flapping wing

**Rahul Kumar** [a], **Devranjan Samanta** [a,*,1] and **Srikant S. Padhee** [a,*,2]

[a] *Department of Mechanical Engineering, Indian Institute of Technology Ropar Rupnagar-140001, Punjab, India*

*Corresponding author:

[1]E-mail: devranjan.samanta@iitrpr.ac.in

[2]E-mail: sspadhee@iitrpr.ac.in

## 7. Pressure distribution on the upper and lower surfaces

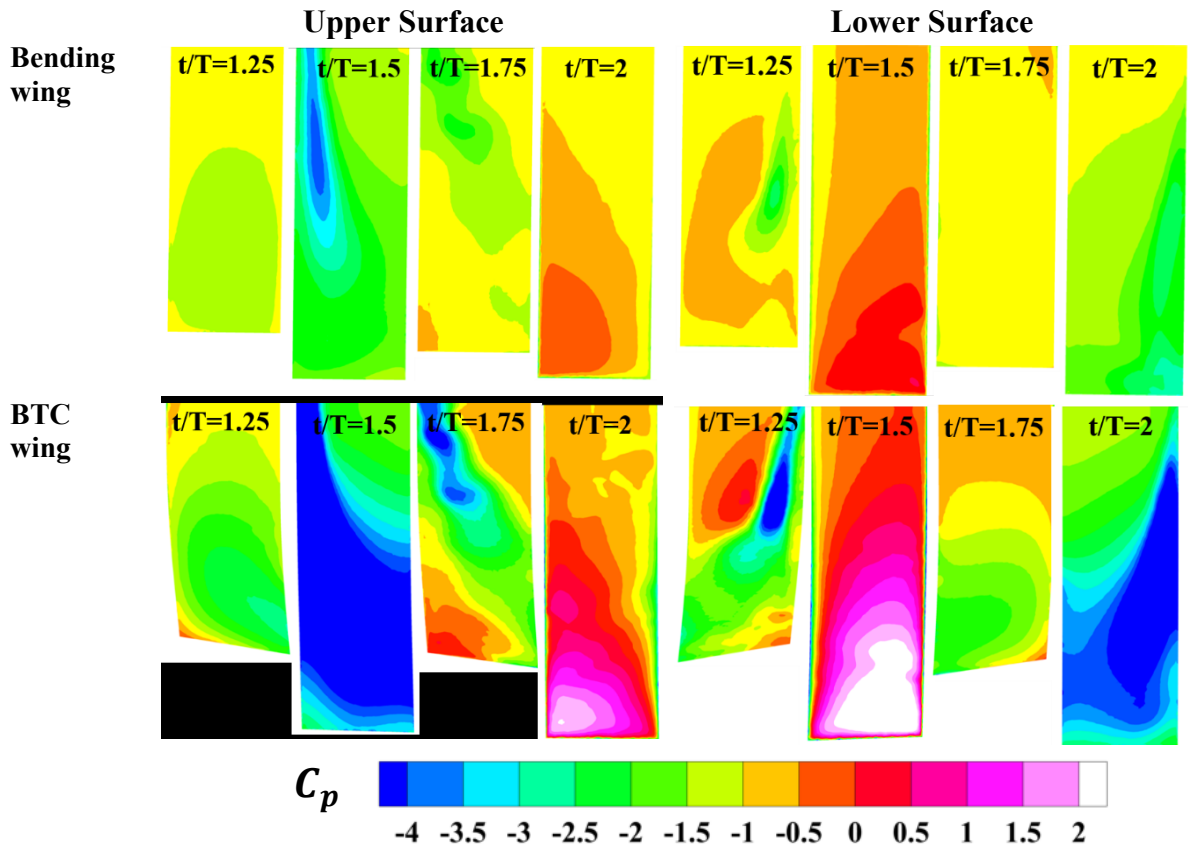

Fig. SI-I. Pressure distribution contours at various times throughout the flapping cycle. **The first row** shows the vorticity contours for the bending wing, while **the second row** displays the vorticity contours for the BTC wing.